\documentclass[aps,amsmath,amssymb,preprintnumbers,groupedaddress,twocolumn]{revtex4-1}
\usepackage{tikz}

\newcommand{\drawsquare}[2]{\hbox{%
\rule{#2pt}{#1pt}\hskip-#2pt%  left vertical
\rule{#1pt}{#2pt}\hskip-#1pt%  lower horizontal
\rule[#1pt]{#1pt}{#2pt}}\rule[#1pt]{#2pt}{#2pt}\hskip-#2pt%  upper horizontal
\rule{#2pt}{#1pt}}% right vertical
\newcommand{\fund}{\raisebox{-.5pt}{\drawsquare{6.5}{0.4}}}%  fund
\newcommand{\Ysymm}{\raisebox{-.5pt}{\drawsquare{6.5}{0.4}}\hskip-0.4pt%
\raisebox{-.5pt}{\drawsquare{6.5}{0.4}}}%  symmetric second rank
\newcommand{\Yasymm}{\raisebox{-3.5pt}{\drawsquare{6.5}{0.4}}\hskip-6.9pt%
\raisebox{3pt}{\drawsquare{6.5}{0.4}}}%  antisymmetric second rank
\newcommand{\ov}{\overline}

\newcommand{\bZ}{\mathbb{Z}}

\begin{document}

\preprint{NSF-KITP-13-214}

\title{Anomaly Nucleation Constrains $SU(2)$ Gauge Theories}
\author{James Halverson} \affiliation{Kavli Institute for Theoretical
  Physics, University of California\\ Santa Barbara, CA 93106-4030 USA}

\begin{abstract}
  We argue for the existence of additional constraints on $SU(2)$
  gauge theories in four dimensions when realized in ultraviolet
  completions admitting an analog of D-brane nucleation. In type II
  string compactifications these constraints are necessary and
  sufficient for the absence of cubic non-abelian anomalies in certain
  nucleated $SU(N>2)$ theories. It is argued that they appear quite
  broadly in the string landscape. Implications for particle physics
  are discussed; most realizations of the standard model in this
  context are inconsistent, unless extra electroweak fermions are
  added.
\end{abstract}

\maketitle

\noindent {\bf I. Introduction}

Despite its many successes, the standard model of particle physics is
an incomplete description of Nature. Some of its shortcomings, such as
the absence of a cold dark matter candidate, can be amended in quantum
field theory; others, such as the absence of quantum gravity, require
a more robust framework for ultraviolet completion.  An important
question is whether constraints on gauge theories in that framework
differ from those of generic quantum field theories, and whether these
differences have implications for low energy physics.

In this letter we focus our attention on $SU(2)$ gauge theories, whose
relevance for Nature cannot be overstated; they govern the weak
interactions \cite{Weinberg:1967tq}, the recently discovered Higgs
boson \cite{Aad:2012tfa,*Chatrchyan:2012ufa}, and perhaps dark matter.
We demonstrate that there are additional constraints on these theories
when realized in ultraviolet completions admitting certain dynamical
processes. These include broad regions of the string landscape.  Most
realizations of the standard model in this context are inconsistent,
unless extra electroweak fermions are added.

The physical argument relies critically on a beautiful property of
string theory.  There, gauge theories are often carried by charged
objects, such as D-branes \cite{Polchinski:1995mt}. If these objects
can pair produce and embed the $SU(2)$ theory into an $SU(N)$ theory,
then the $SU(2)$ theory must satisfy additional constraints necessary
for consistency of the $SU(N)$ theory, but not its own.  D-brane
nucleation is an example of such a process, and we will study the
constraints in this light.  A natural objection is that such a system
is unstable. Indeed this is true, but the gauge theories arising in it
must nevertheless be anomaly free.

String consistency conditions are stronger than those of quantum field
theory. For example, it has been argued \cite{Vafa:2005ui} that there
are effective theories which do not not admit string embeddings, and
also that matter representations are constrained (see
\cite{Grassi:2013kha} for a recent discussion). Here, we obtain a
different type of constraint, placed on one theory to ensure the
consistency of another related by a dynamical process. Perhaps this
basic mechanism could be applied to other transitions in the string
landscape.

This letter is organized as follows.  In section II we present the
physical argument for new constraints.  In section III we derive this
idea in the string landscape. In section IV we discuss implications
for particle physics.

\vspace{2.8cm}
\noindent {\bf II. The Physical Argument}
 
It is already evident in quantum field theory that four-dimensional
$SU(2)$ gauge theories are special within the broader class of $SU(N)$
theories. The latter receive $SU(N)^3$ anomaly contributions from Weyl
fermions in complex representations of the gauge group, and
consistency constrains the allowed representations. However, there is
no such constraint on $SU(2)$ theories, since $SU(2)$ does not have
complex representations. Furthermore, $SU(2)$ theories with an odd
number of Weyl fermion doublets are inconsistent \cite{Witten:1982fp}.
This arises from the fact the $\pi_4(SU(2)) = \bZ_2$, but since
$\pi_4(SU(N>2)) = 0$, there is no corresponding constraint on those
theories.  The former constraints are stronger than the latter.

Our main point can be made in a simple example before turning to
string theoretic realizations. Consider an $SU(2)$ gauge theory in
four dimensions with an even number of left-handed Weyl fermion
doublets. This is a consistent quantum field theory.

Now suppose that this theory is UV-completed into a framework where
gauge theories are carried by charged objects which have dynamics and
can pair produce. If as a result of this process the $SU(2)$ theory
has embedded into a nucleated $SU(N)$ theory, ensuring the absence of
$SU(N)^3$ anomalies can place constraints on the chiral spectrum of
the $SU(2)$ theory. For example, suppose the latter is embedded via
the unhiggsing of an adjoint scalar such that doublets embed either
into the $\fund$ or $\ov \fund$ of $SU(N)$, henceforth $\fund_N$ or
$\ov \fund_N$. The embedding \emph{defines} a way to distinguish two
types of doublets; for notational convenience, denote those of the
first and second type as $\fund_2$ and $\ov \fund_2$, respectively.
Then $SU(N)^3$ anomaly cancellation requires that $\chi(\fund_N)\equiv
\# \fund_N - \# \ov \fund_N$ satisfies
\begin{equation}
0 = \chi(\fund_N) = \chi(\fund_2), 
\end{equation}
where the $SU(2)$ constraint $\chi(\fund_2)=0$ exists due to the
process and ensures the absence of nucleated anomalies.

This is the phenomenon we wish to investigate broadly in the
landscape. In certain corners such constraints have already been
derived and been noted to be stronger than anomaly cancellation; it is
in these corners that we derive the relationship to anomalies in
nucleated theories. In other corners, we utilize nucleation processes
and dualities to argue for their existence. Perhaps explicit
derivations are possible there, as well.

\noindent {\bf III. Traversing the Landscape}

We will begin in type IIa, since there the relationship between the
chiral spectrum of four-dimensional theories and topological
consistency conditions is simple.

\vspace{.2cm}
\noindent {\emph{Large Volume Type IIa Compactifications}}

Consider a large volume type IIa compactification (See
\cite{Blumenhagen:2005mu,*Blumenhagen:2006ci,*Cvetic:2011vz} for
reviews) on a compact Calabi-Yau threefold $X$ with an antiholomorphic
orientifold involution with fixed point locus a three-cycle
$\pi_{O6}\in H_3(X,\bZ)$ wrapped by a spacetime filling O6-plane.
Stacks of $N_a$ spacetime-filling D6-branes which wrap a generic
three-cycle $\pi_a$ and its orientifold image $\pi_a'$ give rise to
$U(N_a)$ gauge theories in four dimensions; the $U(1)\subset U(N_a)$
is often massive, giving $SU(N_a)$ in the infrared.  Chiral matter is
localized at points of D6-brane intersection in $X$; the possible
representations are bifundamentals of two unitary groups or two-index
tensor representations of one.

We would like to study the chiral spectrum of a distinguished D6-brane
stack on $\pi_N$ and its image with $U(N)$ gauge symmetry. Since the
D6-branes and O6-plane carry Ramond-Ramond charge, Gauss' law requires
\begin{equation}
  N(\pi_N + \pi_N')
  + \sum_{a\ne N} N_a (\pi_a + \pi_a') - 4\pi_{O6}=0.
\label{eq:D6 tadpole}
\end{equation}
This is the D6-brane tadpole cancellation condition
\cite{Aldazabal:2000dg}.  The topological intersection numbers of the
branes compute the chiral spectrum as
\begin{align}
\chi(\fund_a,\ov \fund_b)&\equiv \pi_a \cdot \pi_b \qquad  \hspace{.2cm}
\chi(\Yasymm_a)\equiv \frac{1}{2}(\pi_a \cdot \pi_a' + \pi_a \cdot \pi_{O6}) \nonumber \\
\chi(\fund_a, \fund_b)&\equiv \pi_a \cdot \pi_b' \qquad 
\chi(\Ysymm_a)\equiv \frac{1}{2}(\pi_a \cdot \pi_a' - \pi_a \cdot \pi_{O6}), \nonumber 
\end{align} 
Using these and intersecting (\ref{eq:D6 tadpole}) with $\pi_N$ gives
\begin{equation}
  T_N \equiv \chi(\fund_N) + (N-4)\,  \chi(\Yasymm_N) + (N+4) \, \chi(\Ysymm_N) = 0,
\label{eq:IIa spectrum condition} 
\end{equation}
a constraint on the chiral spectrum necessary for D6-brane tadpole
cancellation, and thus global consistency.

This interplay between D6-brane tadpole cancellation and constraints on
the chiral spectrum has been discussed extensively in the type IIa
literature; see \cite{Aldazabal:2000dg,*Uranga:2000xp,*Ibanez:2001nd}
for critical early works. In particular, $T_N=0$ for $N>2$ is the
$SU(N)^3$ anomaly cancellation condition; such anomalies do not exist
for $N=1,2$. In addition, certain $U(1)$ anomalies are cancelled by a
combination of the condition $T_N=0$ and axionic couplings via the
Green-Schwarz mechanism; these include, for example, $U(1)^3$
anomalies for the particular $U(1) \subset U(N)$ \footnote{I am
  grateful to F. Marchesano, G. Shiu, and P. Soler for discussions on
  this point}.  This gives a low energy interpretation of the
constraints $T_2=0$ and $T_1=0$; they play a partial role in
$U(1)$ anomaly cancellation.

We would like to present a different physical understanding of the $T_2$ and
$T_1$ constraints. Though D6-brane charge cancellation in $X$ is
required for consistency, stability is not.  To the system we have
discussed, add a single D6 on $\pi_N$ and a $\ov{D6}$ on a distant but
homologous cycle $\ov \pi_N$ (as well as their orientifold
images). Such a configuration can be reached (with energy cost) by
nucleating a $D6$-$\ov{D6}$ pair on $\ov \pi_N$ and its image, and
then unhiggsing the adjoint scalar associated to the combined D6-brane
system.

Though there is a force between the brane anti-brane pair and they
will annihilate via open string tachyon condensation
\cite{Sen:1998sm}, the worldvolume gauge theories of the D-branes must
nevertheless be anomaly free prior to annihilation; in particular, the
$U(N+1)$ theory on $\pi_N$ must not have $SU(N+1)^3$
anomalies. Similar ideas regarding anomaly cancellation after brane
nucleation have been studied in ten- and six-dimensional theories
\cite{Schwarz:2001sf}.

This relationship between the $T_2$ (or $T_1$) constraint and
nucleated anomalies can be derived.  After adding this pair, there is
a $U(N+1)$ theory on $\pi_N$ and its image, a $U(1)$ theory on the
$\ov{D6}$ on $\ov \pi_N$ and its image, and the $U(N_a)$ theories are
left untouched. Quantitatively, we have added zero (in homology) to
$\eqref{eq:D6 tadpole}$, which now reads
\begin{align}
  (N+1)(\pi_N + \pi_N')
  + \sum_{a\ne N} N_a (\pi_a + \pi_a') - 4\pi_{O6}\nonumber \\ 
  - \ov \pi_N - \ov \pi_N'=0.  
\label{eq:nucleated D6 tadpole}
\end{align}
Intersecting $\pi_N$ with this equation, the terms in the first line
gives the contribution to $SU(N+1)^3$ anomalies from chiral fermions
localized at $D6$-$D6$ intersections, whereas $-\pi_N \cdot \ov
\pi_N=0$, but $-\pi_N \cdot \ov \pi_N'$ can be non-zero. The latter
counts chiral fermions localized at these $D6$-$\ov{D6}$
intersections, and the relative sign is important since the GSO
projection in this sector projects out the opposite chirality fermion.
In all, this calculation gives $T_{N+1}=0$, with the subtlety that
$\chi(\fund_{N+1})$ receives contributions from both $D6$-$D6$ and
$D6$-$\ov{D6}$ intersections.

In summary, the computation before and after nucleation give $T_N=0$
and $T_{N+1}=0$, respectively.  This can be iterated $M$ times, giving
the relationship between the constraint in the setups without and with
$M$ $\ov{D6}$-branes,
\begin{equation}
T_N = 0 \leftrightarrow T_{N+M}=0.
\end{equation}
This immediately gives a simple understanding of the $T_2$ and $T_1$
conditions: they are necessary and sufficient for the absence of
$SU(N+M)^3$ anomalies in this nucleated $U(N+M)$ theory with $N+M >
2$. This should be contrasted with their role in $U(1)$ anomaly
cancellation; since there axionic terms are also required, they are
necessary but not sufficient for $U(1)$ anomaly cancellation.

\vspace{.2cm}
\noindent {\emph{More Derivations in the Landscape}}

We would like to discuss two more derivations of the constraints in
the landscape; one in type IIb, and the other in type I and their
heterotic $SO(32)$ duals.

Consider large volume type IIb flux compactifications with
intersecting stacks of D7-branes carrying abelian worldvolume fluxes
$F_a$. The brane stacks wrap divisors $D_a$ of a Calabi-Yau threefold
$X$ and carry $U(N_a)$ gauge symmetry.  For brevity, consider the case
without orientifolds and a distinguished $U(N)$ theory, this time with
a $D7$-brane stack on $D_N$ with flux $F_N$. The D7 and D5 tadpole
cancellation conditions read
\begin{align}
N\, D_N + \sum_{a\ne N} N_a\, D_a = 0 \nonumber \\ N\, D_N \wedge F_N + \sum_{a\ne N} N_a\, D_a\wedge F_a = 0,
\label{eq:D7 tadpole}
\end{align}
respectively, with Poincar\' e duality implied.  Wedging the D7
tadpole with $F_N \wedge D_N$ and the D5 tadpole with $D_N$ gives
$\sum N_a D_a \wedge D_N \wedge (F_N-F_A)=0$. Rewriting in terms of
the spectrum, this gives $\sum_a N_a\, \chi(\fund_N,\ov \fund_a) =
\chi(\fund_N) = 0$, a constraint necessary for D7 and D5 tadpole
cancellation. Now consider a brane nucleated system with a D7 on $D_N$
and a $\ov {D7}$ on a distant but homologous divisor $\ov D_N$ with
worldvolume fluxes $F_N$ and $\ov F_N$ in the same cohomology
class. This adds zero in cohomology to the D7 and D5 tadpole condition
but gives the constraint $\chi(\fund_{N+1}) = 0$.  Taking $N=2$, the
constraint on the $U(2)$ theory is necessary and sufficient for the
absence of cubic non-abelian anomalies in the nucleated theory.

We see that $SU(N>2)^3$ anomaly cancellation and the additional
$T_2=0$ and $T_1=0$ constraints are a consequence of D7 and D5 tadpole
cancellation.  Since background three-form fluxes contribute to the D3
tadpole, they can be added without spoiling these chiral spectrum
constraints.  These are the fluxes critical in the popular moduli
stabilization scenarios \cite{Kachru:2003aw,*Balasubramanian:2005zx}
which give most of the known landscape of string vacua. Thus, the
constraints appear broadly in the known landscape.

As a final example, consider the type I or $SO(32)$ heterotic string
compactified to four dimensions on a Calabi-Yau threefold $X$ endowed
with a holomorphic vector bundle $V=\bigoplus_{m={K+1}}^{K+L} V_m
\oplus \bigoplus_{i=1}^K L_i$, where the structure group of $V_m$ and
$L_i$ are $U(N_m)$ and $U(1)$, respectively. This is more generic than
the common ansatz of $V$ with $SU(N)$ structure group. The
four-dimensional gauge algebra has an $SU(N_i)$ factor, and in
\cite{Blumenhagen:2005pm} it was shown that the $SU(N_i)^3$ anomaly is
$A_{SU(N_i)^3} \sim 2 \int c_1(L_i) \times \text{Tad}$, where Tad is a
four-form expression which must be cohomologically trivial for
consistency via D5 tadpole cancellation and the B-field Bianchi
identity in the type I and heterotic string, respectively.

Again, $SU(N_i)^3$ anomaly cancellation is ensured by a topological
consistency condition and there is also a constraint for $N_i=2$.  It
is interesting that such a constraint exists in the heterotic string,
since D-brane nucleation has played major a role thus far and does not exist
in the heterotic string. Via nucleation of magnetized D9-branes in the
type I case the $SU(2)$ constraints can likely be related to nucleated
$SU(N_i)^3$ anomaly cancellation. Doing so in a way consistent with
D7-brane tadpole cancellation likely requires introducing an
instability in the gauge bundle, pointing to a heterotic
interpretation.

\vspace{.2cm}
\noindent {\emph{Existence Arguments}}

Though (to our knowledge) similar constraints have not been explicitly
derived in four-dimensional compactifications of M-theory, F-theory,
and the heterotic $E_8\times E_8$ superstring, we would like to
present arguments in favor of their existence, utilizing the existence
of nucleation-type processes and string dualities.

We began by discussing type IIa intersecting D6-brane
compactifications. If supersymmetric, these lift to compactifications
of M-theory on seven manifolds with $G_2$ holonomy, in which case
vector (chiral) multiplet data is captured by codimension four (seven)
singularities in the geometry. An important question is whether the
$D6$-$\ov{D6}$ annihilation process critical to the physics of the
constraints in IIa has a known M-theory lift, even locally. Such
gravitational solutions have in fact been constructed
\cite{Sen:1997pr,*Rabadan:2002zq}; prior to annihilation the system is
described by a bolt singularity, and afterwards a Taub-NUT.  If such a
process relates an $SU(2)$ and $SU(N)$ theory it is plausible that
$SU(N)^3$ anomaly cancellation can be used to constrain the $SU(2)$
theory obtained after annihilation.
 
The T-dual type IIb picture with D7-branes lifts to F-theory.  In its
weak coupling limit, F-theory is simply a geometrization of type IIb.
$D7$-${\ov{D7}}$ nucleation modifies the IIb axiodilaton profile, and
to the author's knowledge the geometric F-theory lift is not known; it
certainly must modify the geometry significantly, as the presence of
the new seven-branes introduces new moduli.  Perhaps the appropriate
modification of the geometry can be extended outside of the weakly
coupled regime; if so, consistency requires the absence of nucleated
anomalies.  As further evidence, it is known \cite{Cvetic:2012xn} that
$SU(N)^3$ anomalies are automatically cancelled in $d=4$ F-theory
compactifications with appropriately specified ``$G_4$-flux,'' and the
mechanism is equivalent to D7 and D5 tadpole cancellation in the weak
coupling limit, which were critical above.

Finally, consider the $E_8 \times E_8$ heterotic string on a
Calabi-Yau threefold. Suppose the compactification has an F-theory
dual; then vector bundle moduli which Higgs $E_8\times E_8$ to the
four-dimensional gauge group map to complex structure moduli in
F-theory, which determine the Higgsing of two $E_8$ seven-branes via
unfolding.  The additional seven-brane moduli associated to the
nucleation process suggested in the last paragraph would require
passing to a different heterotic vector bundle with more moduli; since
supersymmetry would be broken via the nucleation process, the new
bundle must be unstable.  This may provide an avenue for an explicit
derivation.

\vspace{.2cm}
\noindent {\emph{Symplectic Realizations of $SU(2)$}}

We would like to note another possibility, where an $SU(2)$ gauge
theory is realized as $Sp(1)$. In known cases, there are not
constraints analogous to $T_2=0$; e.g. in type IIa there are not
constraints on their chiral spectra necessary for D6 tadpole
cancellation. This matches nicely with the fact that such theories
would nucleate $Sp(N)$ rather than $U(N)$ theories, which do not have
cubic non-abelian anomalies. See \cite{Cremades:2003qj} for a IIa discussion.

Realizing $Sp(N)$ theories can require
additional geometric constraints; e.g. in IIa, D6-brane three-cycles
must be orientifold invariant, and in F-theory codimension two
singular loci must induce an automorphism of codimension one
fibers.  Additionally, conventional grand unification is difficult
when $SU(2)_L$ is realized as $Sp(1)$.

\vspace{.5cm}
\noindent {\bf IV. Implications for Particle Physics}

Given the necessity of ultraviolet completion, it is important to
study the potential implications of these constraints for physics
beyond the standard model.

Model-building from the bottom-up in this context \footnote{That is, when
such constraints hold; when $SU(2)$ is realized
as $Sp(1)$, for example, this is not the case.}, the standard model
(or MSSM) itself is incomplete: one must specify more input data,
labeling $SU(2)_L$ doublets as $\fund_2$ or $\ov \fund_2$ according to
their embedding into the nucleated theory.  Our scheme for counting is
as follows: for each set of $F$ fermion doublets with the same
standard model quantum numbers, consider all possible tuples $(\#
\fund_2, \# \ov \fund_2)$ such that the sum is $F$.  Then the three
families of quark and lepton doublets split as $(3,0)$, $(2,1)$,
$(1,2)$ or $(0,3)$. These contribute to $T_2$ as
\begin{equation}
T_2^l\in\{\pm 1, \pm 3\} \qquad \text{and} \qquad T_2^Q \in \{\pm 3, \pm 9 \},
\end{equation}
where there is a factor of $3$ for color in $T_2^Q$. In all, this
gives $16$ possibilities for $T_2^{SM} \equiv T_2^l + T_2^Q$, only two
of which satisfy $T_2^{SM} = 0$.  In the MSSM, the Higgsinos split as
$(1,0)$ or $(0,1)$, contributing
\begin{equation} 
  T_2^{\tilde h_u} \in \{\pm 1\} \,\,\, \text{and} \,\,\,  T_2^{\tilde h_d} \in \{\pm 1\}
\end{equation} 
with only $6$ of the $64$ possibilities being consistent.  

Most standard model and MSSM configurations do not satisfy the
additional constraint on the $SU(2)$ spectrum.  Without probabilistic
information about the likelihood of realizing one configuration over
another in the landscape, it is difficult to draw definitive
conclusions.

Conservatively, though, there are only two possibilities that should
be considered.

\vspace{.2cm}
\noindent {\bf 1)} 
If $T_2^{SM} = 0$, then the quark doublet sector exhibits a (2+1)
family non-universality \footnote{For brevity, we have not considered
  the possibility that $e_l^c$ or $\nu_l^c$ is realized as an
  antisymmetric tensor; in that analysis the family non-universality
  can be relaxed. Adjoint breaking of $SU(5)$ GUTs yields such
  configurations.}.

\vspace{.2cm}
\noindent {\bf 2)} 
If $T_2^{SM} \ne 0$, then new electroweak fermions are required for
the absence of nucleated anomalies.

\vspace{.2cm}
\noindent Identical statements hold when considering $T_2^{MSSM}$.
The quark doublet non-universality could have further implications;
for example, in type II realizations with $U(2)$ gauge symmetry the
diagonal $U(1)$ will forbid some quark Yukawa couplings in string
perturbation theory.

The second possibility is striking: it provides a new theoretical
motivation for exotic electroweak fermions.  In many cases (see
\cite{Anastasopoulos:2006da,*Cvetic:2011iq} for systematic studies in type II) these new
states are vector-like with respect to the standard model but chiral
under another symmetry, in which case they give smaller corrections to
precision electroweak observables than do chiral exotics, but
nevertheless have protected mass.  If protected from decay, the
neutral components of the exotics are excellent WIMP dark matter
candidates; see \cite{Cvetic:2012kj} for a broad discussion in type
II. These exotic particles could be discovered at LHC or in direct
detection experiments in the near future.

Note that since the $T_3=0$ condition is just the $SU(3)^3$ anomaly
cancellation condition, there is not a similar motivation for new
colored fermions.  Thus, the additional constraints motivate exotics
which are \emph{not} in complete GUT multiplets. One might naively
think that such exotics would ruin gauge coupling unification (GCU),
and in fact they will if added as chiral supermultiplets to the
MSSM. However, in theories without weak scale supersymmetry grand
unification could be a virtue of these exotic $SU(2)_L$ states; see
\cite{Barger:2007qb} for a broad treatment. For example, pairs of
exotic $(1,2)_{\pm 1/2}$ fermions have the quantum numbers of
Higgsinos and can improve GCU; this is the minimal extension
\cite{ArkaniHamed:2005yv,*Mahbubani:2005pt} of the standard model
giving dark matter and grand unification.

\vspace{.5cm}
\noindent {\bf Acknowledgments.} It is a pleasure to thank F. Chen,
A. Collinucci, L. Everett, T. Hartman, G. Kane, A. Puhm and especially
D.R. Morrison, A. Pierce, and J. Polchinski for useful
conversations. I am deeply grateful for conversations and
collaborations on related topics with M. Cveti\v{c}, P. Langacker,
H. Piragua, and R. Richter, and for the support and encouragement of
J.L. Halverson. This research was supported by the National Science
Foundation under Grant No. PHY11-25915.

\bibliography{refs}

\end{document}